%% file: main_revised.tex
\documentclass[conference]{IEEEtran}
\IEEEoverridecommandlockouts
\usepackage{cite}
\usepackage{amsmath,amssymb,amsfonts}
\usepackage{algorithmic}
\usepackage{graphicx}
\usepackage{textcomp}
\usepackage{xcolor}
\usepackage{comment}
\usepackage{pgfplots}
\usepgfplotslibrary{polar}
\usepackage[margin=-5pt, skip=4pt]{caption}
\usepackage{subcaption}
\usepackage{tikz-dimline}
\usepackage{wasysym}
\usepackage{balance}

\usepackage{caption} 
\def\BibTeX{{\rm B\kern-.05em{\sc i\kern-.025em b}\kern-.08em
    T\kern-.1667em\lower.7ex\hbox{E}\kern-.125emX}}

\definecolor{matlab1}{rgb}{0,0.4470,0.7410}
\definecolor{matlab2}{rgb}{0.8500,0.3250,0.0980}
\definecolor{matlab3}{rgb}{0.9290,0.6940,0.1250}
\definecolor{matlab4}{rgb}{0.4940,0.1840,0.5560}
\definecolor{matlab5}{rgb}{0.4660,0.6740,0.1880}
\definecolor{matlab6}{rgb}{0.3010,0.7450,0.9330}
\definecolor{matlab7}{rgb}{0.6350,0.0780,0.1840}

\newcommand{\red}[1]{{\color{black}#1}}
    
\begin{document}

\newcommand{\thetas}{\theta_{\mathrm{s}}}
\newcommand{\thetai}{\theta_{\mathrm{i}}}
\newcommand{\Rmat}{\mathbf{R}}

\title{Vehicle-Mounted Steerable Wideband Antenna Array for NATO Band III Tactical Radio Links%
}

\author{Ida Pääkkölä, Mikko Heino, and Taneli Riihonen\\
Faculty of Information Technology and Communication Sciences, Tampere University, Finland\\
email: \texttt{mikko.heino@tuni.fi}}

\maketitle

\begin{abstract}
In this paper, a very compact 2 x 2 wideband beamsteering array with monocone antennas is designed and validated for NATO Band III (1350--2400 MHz) tactical communications with interference suppressing ability. The small 4 x 4 x 4 cm\textsuperscript{3} size of the array enables placing it for example on vehicles with limited space. The array is manufactured and measured, and the interference suppressing capability thereof is studied with a minimum variance distortionless response beamformer and a phase-only beamformer in the presence of a single interferer. It is shown that the designed antenna array can significantly improve the signal-to-noise ratio over 10~dB for angles separated over 15$^\circ$ from the signal-of-interest, and over 20~dB for separation larger than 43$^\circ$ across the operational band.
\end{abstract}


\section{Introduction}

Modern tactical communication systems need to operate in ever-increasing electromagnetically hostile environments where adversary EW systems locate and employ strong jamming against the receivers. The interfering signals severely decrease reception quality at the receiver. Typically, the jamming is countered by increasing the transmit power, deploying directive antennas to spatially filter out the interferer with the radiation pattern, or employing wideband anti-jamming waveforms like frequency hopping signals \cite{poisel2011modern}.

To spatially filter out interfering signals effectively, adaptive beamforming techniques like null-steering are needed, as typically the side- and backlobe suppression of fixed directive antennas is not enough. Currently, the most typical use case for interference reducing antennas are Controlled Reception Pattern Antennas (CRPAs), which electronically beamform to null the interfering signals \cite{Byun2015}. They are typically used with GNSS receivers, for examples in UAVs, as the signals are very susceptible to jamming due to the very low power level. Combined with electronic beamforming, the anti-jamming performance with multiple interfering sources can be further enhanced with physical rotation of the antenna array \cite{sun2022anti}. However, in addition to the very narrowband GNSS receivers, there is a lack of wideband compact antenna systems for tactical communications with adaptive interference suppression.

This work studies by simulations and measurements a compact wideband four-element antenna array that could be easily mounted for example vehicles or manpacks with limited space to be used with an adaptive beamformer. Wideband operation across the whole NATO Band III (1350--2400 MHz) is targeted. Specifically, the work combines compact spacing well below 0.5$\lambda$ at the band center and wideband operation with over 50\% bandwidth relative to the center frequency.

\begin{figure}[t]
    \centering
    \includegraphics[width=0.43\textwidth]{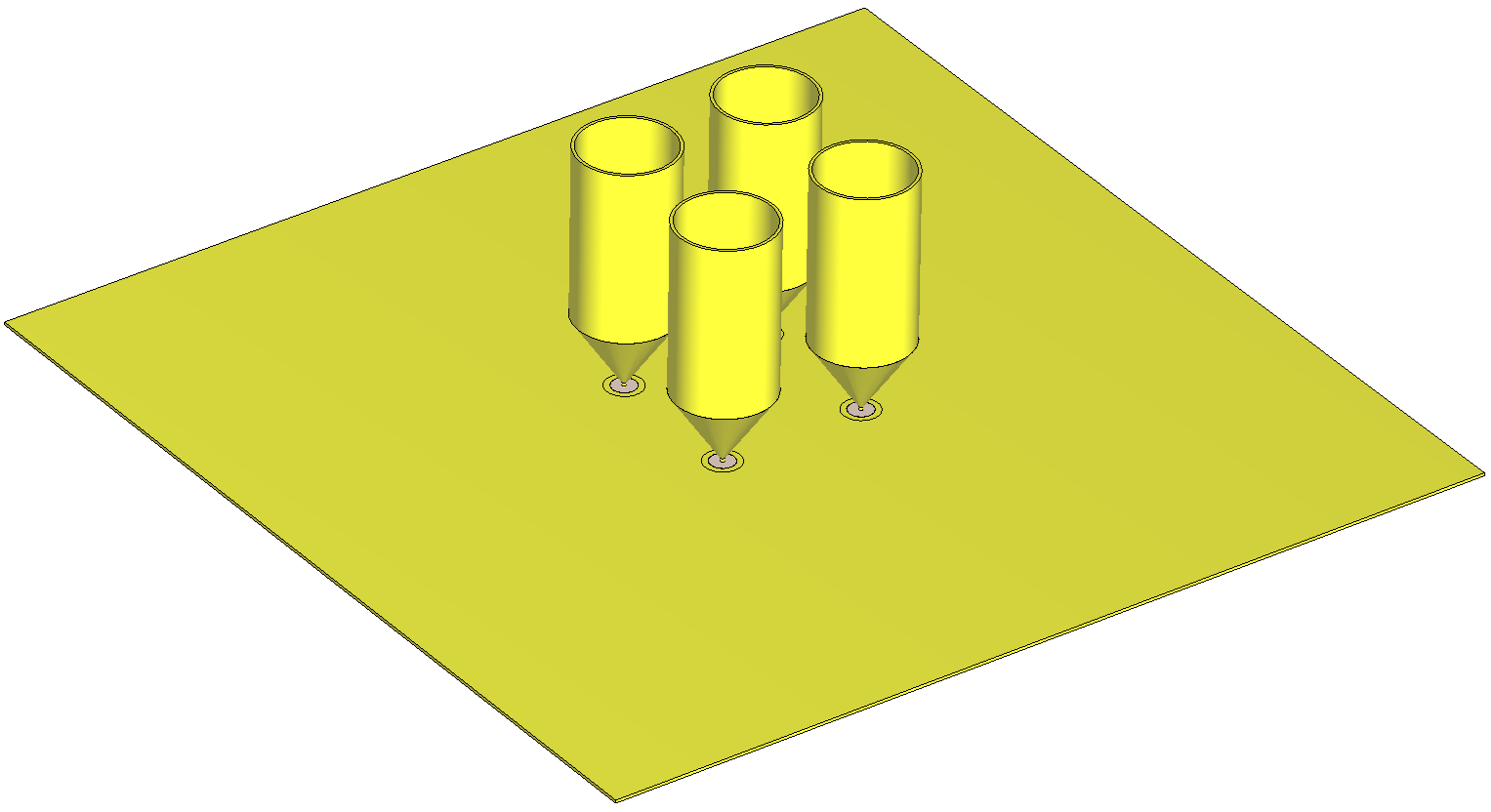}
    \caption{The designed wideband interference suppressing beamsteering array prototype for NATO Band III (1350--2400 MHz).}
    \vspace{-0.9mm}
    \label{fig:designedarray}
\end{figure}

Currently no existing work combines compact spacing ($<$0.5$\lambda$) and wideband ($>$50\% BW) phase-only null-steering. In \cite{Madni2023}, a narrowband 125 mm diameter four-element GNSS array using defected ground structures (DGS) and an absorber is used to enable 48 dB null depth with 6\% of relative BW. In \cite{Iizasa2014}, a 2$\times$2 slot dipole array antenna for beamsteering was presented operating from 2.3 to 2.6 GHz with the size of 16$\times$8 cm\textsuperscript{2}. In \cite{wang2013highly}, a highly compact 2$\times$2 monopole array operating from 2.4 to 2.5 GHz with the size of 1.1$\times$1.1$\times$2.7 cm\textsuperscript{3} and in \cite{ZhaoFourElement2023} a 2$\times$2 monopole array operating from 3.45 to 3.66 GHz with the size of 2$\times$2$\times$2 cm\textsuperscript{3} were presented.

In this work, it is shown that with a minimal 0.15$\lambda$ spacing at the center frequency and very wideband antenna elements, a null-steering beamforming array can be designed operating successfully across the full NATO III Band of 1350 to 2400 MHz without a significant penalty due to mutual coupling. The designed four-element circular/square array with the size of 4$\times$4$\times$4 cm\textsuperscript{3} is manufactured and validated by S-parameter and radiation pattern measurements.

The performance of the antenna array is analyzed with a phase shifter-based and minimum variance distortionless response (MVDR) beamforming algorithms in a scenario of tactical communications being jammed from varying angles relative to the angle of reception of the signal-of-interest. It is shown that even with a phase-only beamformer, the designed antenna array can improve the SINR over 10 dB for angles separated over 15$^\circ$ from the signal-of-interest (SOI), and over 20 dB for separation larger than 43$^\circ$.

\section{Array Design}
\subsection{Antenna Element Selection and Design}

Several antenna element types were evaluated for wideband operation across NATO Band III (1350--2400 MHz): monopole, normal-mode helix, and monocone \cite{McDonald&Filipovic-BandwidthMonoconeAntennas}. With the monopole, only 63\% band coverage could be achieved with \mbox{-10 dB} impedance matching level, which is insufficient for the target frequency range. The normal-mode helix exhibited severe pattern instability across frequencies, making it unsuitable for predictable array operation.

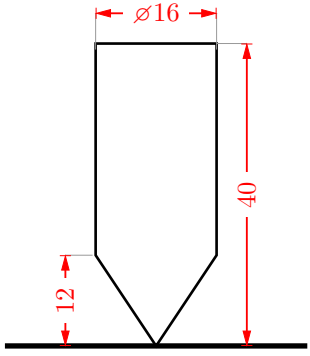
\begin{figure}[t]
\centering
\input{figs/dimensions}
\caption{Monocone element geometry with dimensions in mm.}
\label{fig:monocone}
\end{figure}

The monocone demonstrated superior performance, achieving 94.8\% band coverage with stable omnidirectional patterns across the band. The selected element geometry is shown in Fig.~\ref{fig:monocone}, with base diameter 16 mm, height 40 mm, and cone angle of 33.7 degrees. The simulated and measured \mbox{S-parameters} of the designed monocone antenna placed on a 150 × 150 mm ground plane are shown in Fig. \ref{fig:SparamSingle} with \mbox{-10 dB} impedance matching level obtained from 1.45 GHz up to 2.4~GHz as well as a good match between the simulations and the measurement.

\subsection{Antenna Array Geometry Design}
A 2$\times$2 square (or, equivalently, circular) layout was selected as the minimum configuration for azimuth null steering, providing three degrees of freedom from four antennas.

The antenna element spacing was optimized through full-wave simulations for good null-steering performance across the whole band of 1350 to 2400 MHz. Spacings of 0.20$\lambda_c$ and 0.25$\lambda_c$ defined at the band center frequency 1875 MHz achieved poor null depth at 2400 MHz, while 0.14$\lambda_c$ showed the opposite trend. A spacing of 0.15$\lambda_c$ (i.e., 24 mm at 1875 MHz) was selected as the compromise providing balanced null performance across the band and compact size.

This tight spacing increases the mutual coupling level in the array, but ensures grating-lobe-free operation for the whole band and compact footprint (40$\times$40 mm\textsuperscript{2} array). The array was mounted on a 350$\times$350 mm\textsuperscript{2} ground plane representing a vehicle rooftop. The designed array is shown in Fig. \ref{fig:designedarray}.

\begin{figure}[t]
   \centering
   \includegraphics[width=0.95\columnwidth]{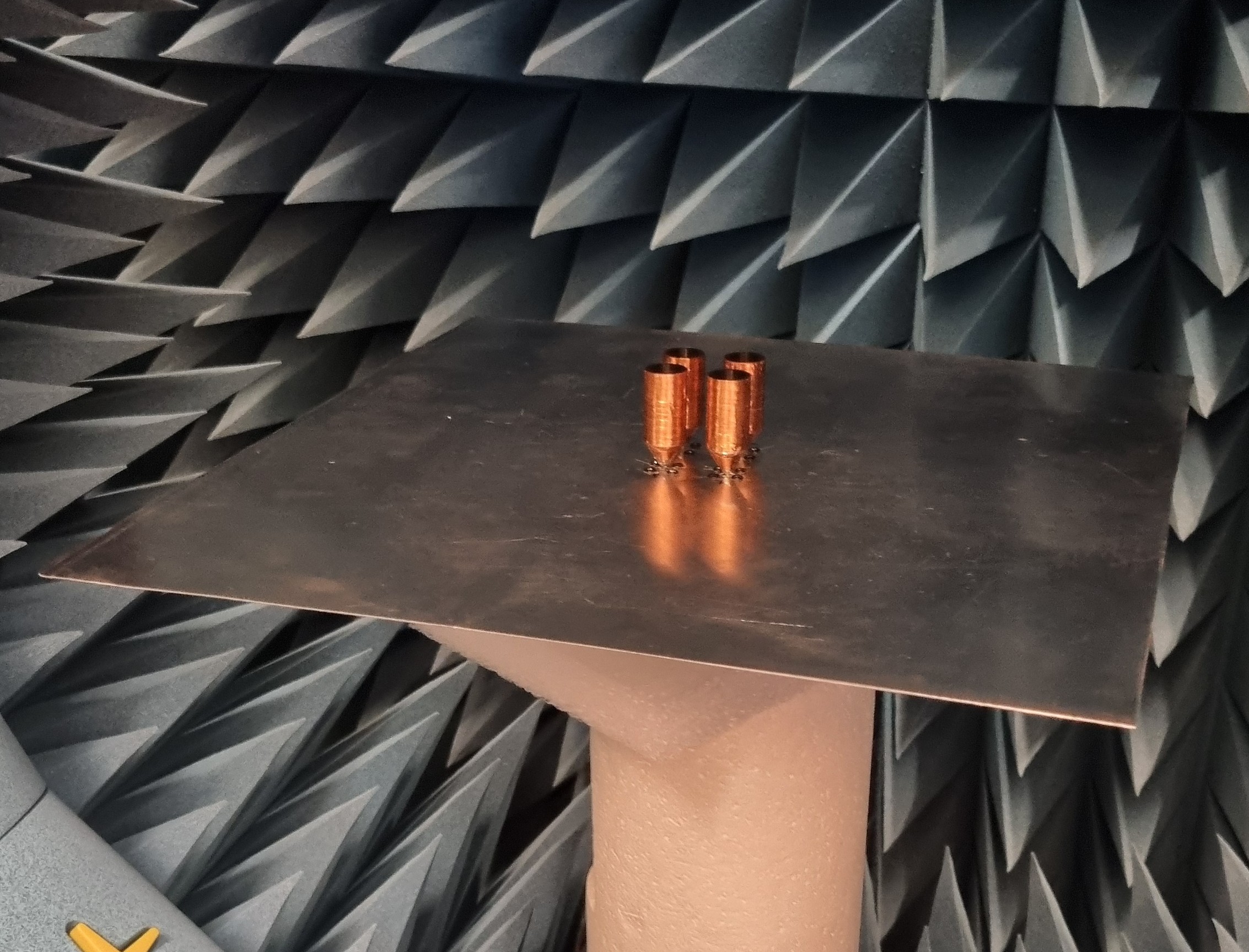}
   \caption{The manufactured antenna array with a 350 mm ground plane in the radiation pattern measurement.}
    \label{fig:manufacturedarray}
\end{figure}

\subsection{Beamforming for Interference Suppression}

The array was designed to be used with adaptive beamsteering methods like phase-only null-steering which could be implemented with simple phase-shifters feeding the array or adaptive beamformers with complex weights like a minimum variance distortionless response (MVDR) beamformer. The MVDR reduces the interference plus noise energy and maximizes the SINR for desired signal without distorting it. It has the benefit of requiring only the knowledge of the desired signal direction of arrival (DOA) to maximize the SINR as opposed to phase-shifter-based null-steering which requires adaptive scanning of the interference angle.

A signal model for the received signals $\mathbf{y}(t)$ for antenna elements in the array is defined as
\begin{equation}
    \mathbf{y}(t) = \sqrt{P_{\mathrm{s}}} \, \mathbf{s}(\thetas) \, x_{\mathrm{s}}(t) + \sqrt{P_{\mathrm{i}}} \, \mathbf{s}(\thetai) \, x_{\mathrm{i}}(t)+\sqrt{P_{\mathrm{n}}}w_\mathrm{n}(t),
\end{equation}
where $x_{\mathrm{s}}(t)$ is the signal-of-interest (SOI), $x_{\mathrm{i}}(t)$ is the interfering signal, $w_{\mathrm{n}}(t)$ the receiver noise, $\thetas$ is the direction of the SOI, $\thetai$ the direction of the interference, $\mathbf{s}(\theta)$ the array steering vector to direction $\theta$, and $\sqrt{P_{\mathrm{s}}}$, $\sqrt{P_{\mathrm{i}}}$ and $\sqrt{P_{\mathrm{n}}}$, the SOI, interference and noise powers, respectively. It is assumed that $x_{\mathrm{s}}(t)$ and $x_{\mathrm{i}}(t)$ are zero mean and uncorrelated.

The covariance matrix $\Rmat$ for the received signals is defined as 
\begin{equation}
    \Rmat = E[\mathbf{y}(t) \mathbf{y}^H(t)] = P_{\mathrm{s}} \, \mathbf{s}(\thetas) \mathbf{s}^H(\thetas) + P_{\mathrm{i}} \, \mathbf{s}(\thetai) \mathbf{s}^H(\thetai)+P_{\mathrm{n}}.
\end{equation}

The optimal weights to maximize the SINR with MVDR are then defined as
\begin{equation}
\mathbf{w_\mathrm{MVDR}} = \frac{\Rmat^{-1}\,\mathbf{s}(\thetas)}{\mathbf{s}^H(\thetas)\,\Rmat^{-1}\,\mathbf{s}(\thetas)}.
\end{equation}
For the phase-only beamformer, the weights are obtained from the optimization problem:
\begin{align}
\max_{\mathbf{w}} \quad 
& \frac{\left| \mathbf{w}^H \mathbf{s}(\thetas) \right|^2}
{\mathbf{w}^H \mathbf{R} \mathbf{w}} \\
\text{s.t.} \quad 
& |\mathrm{w_n}| = \frac{1}{\sqrt{N}},
\end{align}
where $N$ is the number of antenna elements. The optimization is performed with a grid search refined with interior-point method. The received signal after the beamformer is then obtained as 
\begin{align}
y(t) = \mathbf{w}^H \mathbf{y}(t).
\end{align}
% Saisko tähän kohti kuvauksen, miten optimointiongelma on ratkaistu tuloksissa?

\section{Array Measurements}
To characterize the array for beamforming, the embedded element pattern (EEP) approach was employed, where each element was measured individually with remaining elements terminated with 50~$\Omega$ loads. This captured realistic mutual coupling effects that dominate array behavior at 0.15$\lambda$ spacing. All four element patterns were measured across the whole band of operation. Spherical near-field measurements of the antenna array were performed using an MVG Starlab system with the far-field obtained through a mathematical transform. The manufactured array during the measurements is shown in Fig. \ref{fig:manufacturedarray}.

\begin{figure}
    \centering
    \input{figs/portpatterns.tex}
    \vspace{0.2cm}
    \caption{The measured realized gain of individual element patterns in the azimuth plane at 1870 MHz.}
    \label{fig:portpatternsazimuth}
\end{figure}
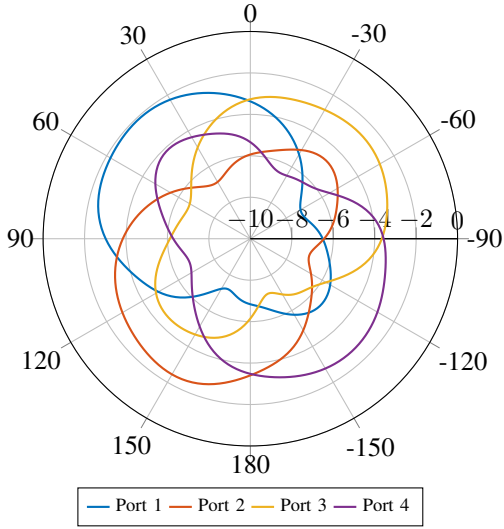

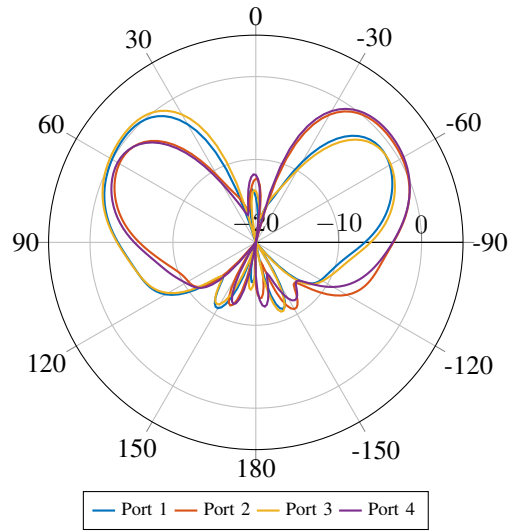
\begin{figure}
    \centering
    \input{figs/portpatternselevation.tex}
    \vspace{0.2cm}
    \caption{The measured realized gain of individual element patterns in the elevation plane at 1870 MHz.}
    \label{fig:portpatternselevation}
\end{figure}

The individual measured patterns of the antenna elements in the array are shown in Figs. \ref{fig:portpatternsazimuth} and \ref{fig:portpatternselevation} as realized gain plots. It is seen in Fig. \ref{fig:portpatternsazimuth} that the patterns are omnidirectional across the band with 2 dB more gain outward from the array due to close proximity of the other elements. In the elevation plane in Fig. \ref{fig:portpatternselevation}, the ground plane directs the main beam up from the horizontal plane so the maximum gain is obtained at around \red{40$^\circ$ to 65$^\circ$} elevation angle. 

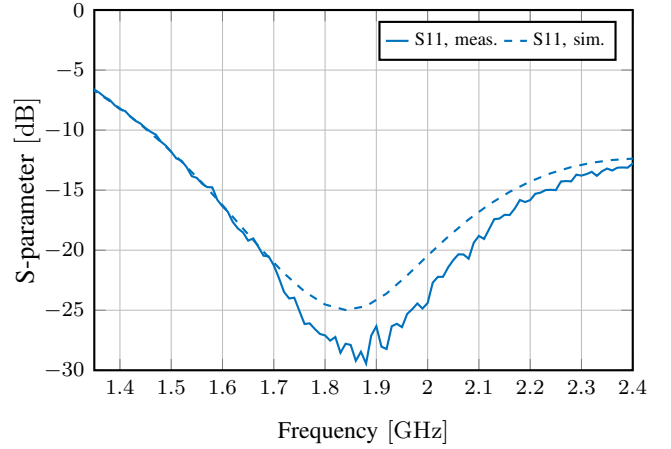
\begin{figure}
    \centering
    \input{figs/sparameterssingle.tex}
    \caption{Measured and simulated S-parameters of a single element in the array.}
    \label{fig:SparamSingle}
\end{figure}

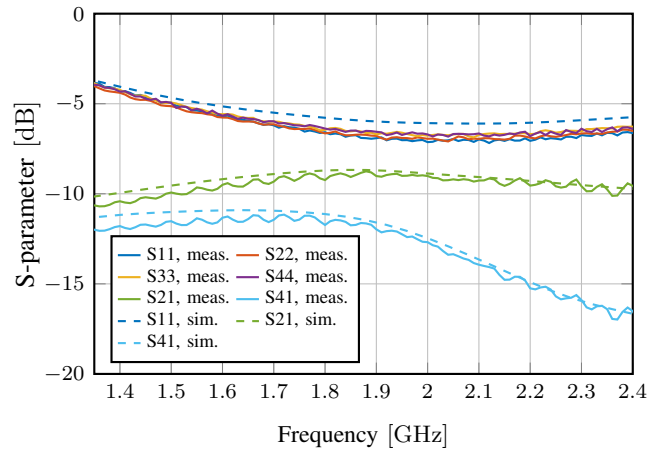
\begin{figure}
    \centering
    \input{figs/sparameters.tex}
    \caption{Measured and simulated S-parameters of the manufactured prototype array.}
    \label{fig:arraysparameters}
\end{figure}

The measured S-parameters for the array elements are shown in Fig. \ref{fig:arraysparameters}. It is seen that the close proximity of the antennas in the array detunes the matching of the elements to a level of about -6 dB as opposed to the below -10 dB matching level of a single \red{isolated} element in Fig. \ref{fig:SparamSingle}. \red{This coupling between the elements} causes a decrease of 3 to 4 dB in the total efficiency. However, the coupling between adjacent elements is at the level of -9 dB. The antennas could be further tuned to avoid the detuning due to the close proximity of the other elements and increase the efficiency.

\section{Beamsteering Results}

The measured complex radiation patterns of each antenna element in the array are used as steering vectors in the beamforming analysis. The beamforming algorithms with the measured array are studied in a scenario where the signal-of-interest is arriving to the array from directions 0$^\circ$ or 45$^\circ$ degree relative to the array and the direction of one interfering signal is swept across the whole azimuth plane. The SNR for the incoming signal is defined as $\mathrm{SNR}=\frac{P_{\mathrm{s}}}{P_{\mathrm{n}}}$ and the INR of the interfering signal as $\mathrm{INR}=\frac{P_{\mathrm{i}}}{P_{\mathrm{n}}}$ relative to the receiver noise level. When defining the SNR and INR of the signals, an isotropic antenna is assumed at the receiver. Then after the adaptive beamforming with the real antenna patterns, the real SINR at the receiver is obtained depending on how well the beamformer is able to cancel the interference. In ideal case, the interference is cancelled under the noise floor of the receiver and the SINR is defined solely by the power of the signal-of-interest.

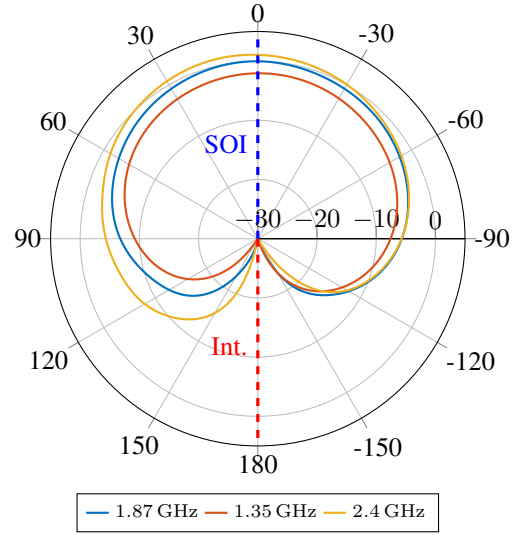
\begin{figure}
    \centering
    \input{figs/patterns.tex}
    \vspace{0.0cm}
    \caption{The realized gain pattern of the array with measured steering vectors with an MVDR beamformer with SOI angle 0$^\circ$ and interferer angle 180$^\circ$.}
    \label{fig:MVDRpatternINT180}
\end{figure}

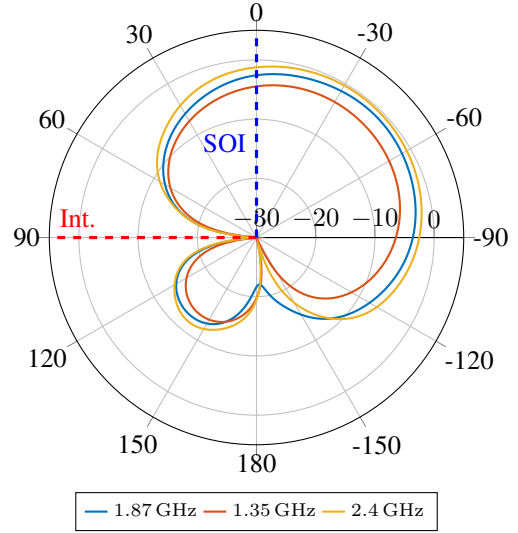
\begin{figure}
    \centering
    \input{figs/patternsinterferer90.tex}
    \vspace{0.0cm}
    \caption{The realized gain pattern of the array with an MVDR beamformer with SOI angle 0$^\circ$ and interferer angle 90$^\circ$.}
    \label{fig:MVDRpatternINT90}
\end{figure}

Figs.~\ref{fig:MVDRpatternINT180} and \ref{fig:MVDRpatternINT90} show the array patterns across the full band when using the MVDR beamformer to cancel interference from azimuth directions 180$^\circ$ and 90$^\circ$. It is seen that the array is able to steer the null both in the band low- and high-frequency limits. 

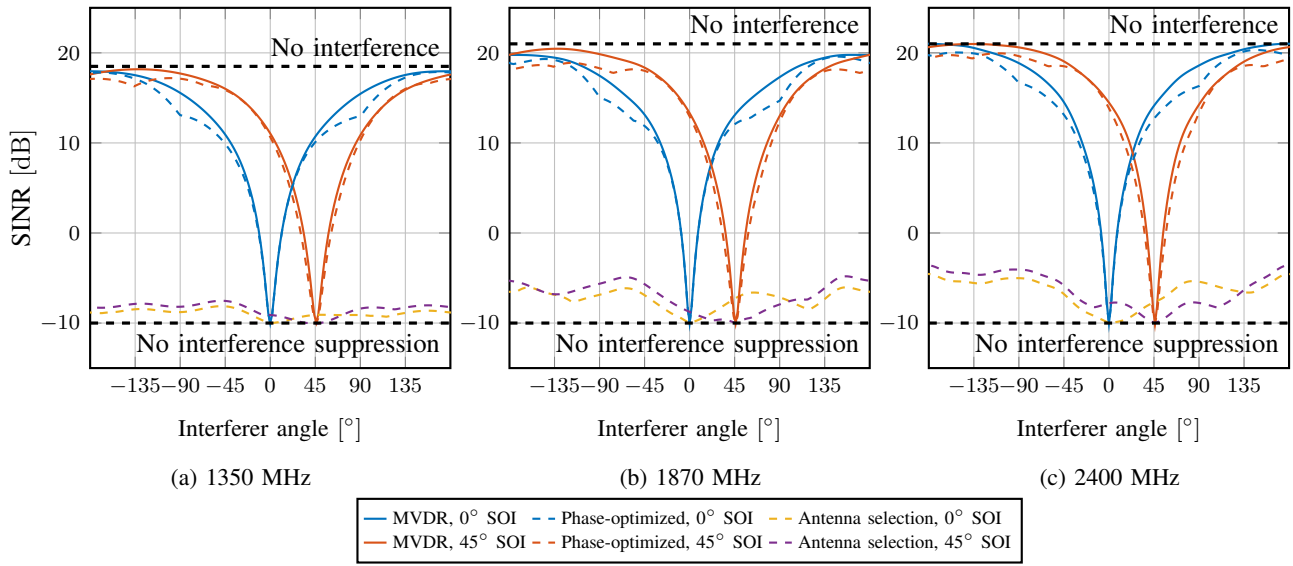
\begin{figure*}[t]
    \centering
    \begin{subfigure}[b]{0.35\textwidth}
         \input{figs/beamsteeringSINR1350_v2.tex}
         \caption{1350 MHz}
     \end{subfigure}\hspace*{-0.4cm}%
     \begin{subfigure}[b]{0.35\textwidth}
         \input{figs/beamsteeringSINR_v2.tex}
         \caption{1870 MHz}
     \end{subfigure}\hspace*{-0.8cm}%
     \begin{subfigure}[b]{0.35\textwidth}
         \input{figs/beamsteeringSINR2400_v2.tex}
         \caption{2400 MHz}
     \end{subfigure}
    \vspace{1cm}
    \captionsetup{width=\textwidth}
    \caption{\red{SINR obtainable with the measured prototype array by beamsteering or antenna selection with varying interferer angle and signal-of-interest coming from either 0° or 45° angle.}}
    \label{fig:interferencesweep}
\end{figure*}

Fig.~\ref{fig:interferencesweep} shows the obtained SINR when the interferer angle is swept relative to the direction of the signal-of-interest. The SNR for the signal-of-interest is 20 dB and the INR for the interfering signal is 30 dB, i.e., the jamming-to-signal ratio is 10 dB. Two different angles for the SOI are studied based on the geometry of the array, a signal arriving from 0$^\circ$, the normal direction of the array, and from 45$^\circ$, diagonally relative to the antenna elements. 

\red{The SINR obtained with antenna selection method (for reference), MVDR beamformer and a phase-only beamformer are plotted.} Two horizontal lines are plotted, one representing what would be the SINR without any interfering signal with the beam steered towards the SOI and one horizontal line when there is no interference suppression at all. \red{In the antenna selection method, only the single antenna with the best SINR is used for reception, i.e., indicating the performance with a switch instead of a beamformer.}

It is seen that for both SOI angles, the MVDR beamformer is able to improve the SINR over 10 dB for angles separated over 15$^\circ$ from the signal-of-interest (SOI) and over 20 dB for separation larger than 43$^\circ$ for frequencies from 1350 to 2400 MHz. When the interferer is separated over 90$^\circ$ from the SOI, the array is able to suppress the interference with over 25 dB improvement and the noise floor of the receiver is reached behind the array at 180$^\circ$. When the interference is coming from the same direction as the signal-of-interest, there is naturally no interference suppression. \red{With the antenna selection method, no significant interference suppression is obtainable because the antenna elements are omnidirectional.}

When changing to the phase-only beamformer, almost as good performance is obtained for the SINR improvement. The decrease in performance is at most 1 to 2 dB from the ideal performance with the MVDR beamformer. This enables easier implementation of the beamformer for example with off-the-shelf phase shifters.

\section{Conclusions}
In this paper, a wideband very compact four-element antenna array was presented operating at the NATO Band III (1350 to 2400 MHz) with interference nulling capability for tactical communications. The small 4$\times$4$\times$4 cm\textsuperscript{3} size of the array enables the array to be easily placed for example on vehicles with limited space. The designed array is manufactured and verified with radiation pattern measurements. It is shown that the array is able to improve the SINR over 10 dB for angles separated over 15$^\circ$ from the signal-of-interest (SOI), and over 20 dB for separation larger than 43$^\circ$ using both MVDR and phase-only beamformers for the whole band. For angles larger than 100$^\circ$, the array is able to cancel the interferer down to the noise floor of a receiver. Further work includes tuning the antenna array to reduce the impedance detuning caused by the close proximity of the elements and shaping the ground plane to increase the gain of the array in the elevation plane towards the desired signal-of-interest.

%\section*{Acknowledgment}

\vspace{0.5cm}
\balance
\bibliographystyle{IEEEtran} % We choose the "plain" reference style
\bibliography{references}

\end{document}

%% file: figs/dimensions.tex
\begin{tikzpicture}[scale=0.1]

\draw[line width=1pt] (0,0) -- (-8,12) -- (-8,40) -- (8,40) -- (8,12) -- cycle;

\draw[line width=2pt] (-20,0) -- (20,0);

\dimline[line style = {red, line width=0.7},
extension start length=-0.0,
extension end length=-0.1] {(12,0)}{(12,40)}{$40$};

\dimline[line style = {red, line width=0.7},
extension start length=0.00,
extension end length=0.30] {(-12,0)}{(-12,12)}{$12$};

\dimline[line style = {red, line width=0.7},
extension start length=0.3,
extension end length=0.3] {(-8,44)}{(8, 44)}{ $\diameter 16$};

\end{tikzpicture}

%% file: figs/portpatterns.tex
%\centering
    \begin{tikzpicture}
    \begin{polaraxis}[
    scale=0.8,
    ylabel style={xshift=0.1cm,yshift=-0.5cm},
    xlabel style={xshift=0.1cm,yshift=-0.5cm},
    legend style={anchor=north, at={(axis cs:270,2)}},
    legend columns=5,
    legend style={font=\scriptsize},
    legend image post style={scale=0.5},
    ymin=-10,
    ymax=0,
    y coord trafo/.code=\pgfmathparse{#1+10},
    y coord inv trafo/.code=\pgfmathparse{#1-10},
    xticklabels={0,-90,-60,-30,0,30,60,90,120,150,180,-150,-120},
    ]
    \addplot+[no marks,matlab1,thick]  table[x expr=\thisrowno{0}+90, y index=1] {data/Port1PatternAzimuth_1870.txt};
    \addplot+[no marks,matlab2,thick]  table[x expr=\thisrowno{0}+90, y index=1] {data/Port2PatternAzimuth_1870.txt};
    \addplot+[no marks,matlab3,thick]  table[x expr=\thisrowno{0}+90, y index=1] {data/Port3PatternAzimuth_1870.txt};
    \addplot+[no marks,matlab4,thick]  table[x expr=\thisrowno{0}+90, y index=1] {data/Port4PatternAzimuth_1870.txt};
    %\addplot+[no marks,matlab2,thick]  table[x expr=\thisrowno{0}+90, y index=1] {data/MVDRPatternSOI0Int180_1350.txt};
    %\addplot+[no marks,matlab3,thick]  table[x expr=\thisrowno{0}+90, y index=1] {data/MVDRPatternSOI0Int180_2400.txt};
    %addplot+[no marks, thick, red, dashed] coordinates {(-90, 5) (-90, -30) };
    %\addplot+[no marks, dashed, thick, black] coordinates {(-30, -30)};

    \addlegendentry{Port 1}
    \addlegendentry{Port 2}
    \addlegendentry{Port 3}
    \addlegendentry{Port 4}

    \end{polaraxis}
    \end{tikzpicture}

%% file: figs/portpatternselevation.tex
%\centering
    \begin{tikzpicture}
    \begin{polaraxis}[
    scale=0.8,
    ylabel style={xshift=0.1cm,yshift=-0.5cm},
    xlabel style={xshift=0.1cm,yshift=-0.5cm},
    legend style={anchor=north, at={(axis cs:270,10)}},
    legend columns=5,
    legend style={font=\scriptsize},
    legend image post style={scale=0.5},
    ymin=-20,
    ymax=5,
    y coord trafo/.code=\pgfmathparse{#1+20},
    y coord inv trafo/.code=\pgfmathparse{#1-20},
    xticklabels={0,-90,-60,-30,0,30,60,90,120,150,180,-150,-120},
    ]
    \addplot+[no marks,matlab1,thick]  table[x expr=\thisrowno{0}+90, y index=1] {data/Port1PatternElevation_1870.txt};
    \addplot+[no marks,matlab2,thick]  table[x expr=\thisrowno{0}+90, y index=1] {data/Port2PatternElevation_1870.txt};
    \addplot+[no marks,matlab3,thick]  table[x expr=\thisrowno{0}+90, y index=1] {data/Port3PatternElevation_1870.txt};
    \addplot+[no marks,matlab4,thick]  table[x expr=\thisrowno{0}+90, y index=1] {data/Port4PatternElevation_1870.txt};
    %\addplot+[no marks,matlab2,thick]  table[x expr=\thisrowno{0}+90, y index=1] {data/MVDRPatternSOI0Int180_1350.txt};
    %\addplot+[no marks,matlab3,thick]  table[x expr=\thisrowno{0}+90, y index=1] {data/MVDRPatternSOI0Int180_2400.txt};
    %addplot+[no marks, thick, red, dashed] coordinates {(-90, 5) (-90, -30) };
    %\addplot+[no marks, dashed, thick, black] coordinates {(-30, -30)};

    \addlegendentry{Port 1}
    \addlegendentry{Port 2}
    \addlegendentry{Port 3}
    \addlegendentry{Port 4}

    \end{polaraxis}
    \end{tikzpicture}

%% file: figs/sparameterssingle.tex
\begin{tikzpicture}
\begin{axis}[
legend style={font=\scriptsize},
legend columns=2,
legend pos=north east,
legend style={inner sep=2pt},
legend style={row sep=-2pt},
legend cell align={left},
legend image post style={scale=0.5},
small,
xlabel=Frequency $\mathrm{[GHz]}$,
ylabel=S-parameter $\mathrm{[dB]}$,
width=0.48\textwidth,
height=0.35\textwidth,
xmin=1.35,
xmax=2.4,
ymin=-30,
ymax=0,
grid=both,
thick,
ylabel near ticks,
ylabel shift = -0.2cm,
%axis y line*=left,
%ytick distance=10
%legend style={yshift=-1cm}
]
\addplot+[no marks,matlab1]  table[x=freq,y=S11] {data/MeasCone.txt};
\addplot+[no marks,matlab1,dashed]  table[x=freq,y=S11] {data/SimCone.txt};

\addlegendentry{S11, meas.}
\addlegendentry{S11, sim.}

\end{axis}
\end{tikzpicture}

%% file: figs/sparameters.tex
\begin{tikzpicture}
\begin{axis}[
legend style={font=\scriptsize},
legend columns=2,
legend pos=south west,
legend style={inner sep=2pt},
legend style={row sep=-2pt},
legend cell align={left},
legend image post style={scale=0.5},
small,
xlabel=Frequency $\mathrm{[GHz]}$,
ylabel=S-parameter $\mathrm{[dB]}$,
width=0.48\textwidth,
height=0.35\textwidth,
xmin=1.35,
xmax=2.4,
ymin=-20,
ymax=0,
grid=both,
thick,
ylabel near ticks,
ylabel shift = -0.2cm,
%axis y line*=left,
%ytick distance=10
%legend style={yshift=-1cm}
]
\addplot+[no marks,matlab1]  table[x=freq,y=S11] {data/MeasSparams.txt};
\addplot+[no marks,matlab2]  table[x=freq,y=S22] {data/MeasSparams.txt};
\addplot+[no marks,matlab3]  table[x=freq,y=S33] {data/MeasSparams.txt};
\addplot+[no marks,matlab4]  table[x=freq,y=S44] {data/MeasSparams.txt};
\addplot+[no marks,matlab5]  table[x=freq,y=S21] {data/MeasSparams.txt};
\addplot+[no marks,matlab6,solid]  table[x=freq,y=S41] {data/MeasSparams.txt};

\addplot+[no marks,matlab1,dashed]  table[x=freq,y=S11] {data/SimSparams.txt};
\addplot+[no marks,matlab5,dashed]  table[x=freq,y=S21] {data/SimSparams.txt};
\addplot+[no marks,matlab6,dashed]  table[x=freq,y=S41] {data/SimSparams.txt};

\addlegendentry{S11, meas.}
\addlegendentry{S22, meas.}
\addlegendentry{S33, meas.}
\addlegendentry{S44, meas.}
\addlegendentry{S21, meas.}
\addlegendentry{S41, meas.}

\addlegendentry{S11, sim.}
\addlegendentry{S21, sim.}
\addlegendentry{S41, sim.}

\end{axis}
\end{tikzpicture}

%% file: figs/patterns.tex
%\centering
    \begin{tikzpicture}
    \begin{polaraxis}[
    scale=0.8,
    ylabel style={xshift=0.1cm,yshift=-0.5cm},
    xlabel style={xshift=0.1cm,yshift=-0.5cm},
    legend style={anchor=north, at={(axis cs:270,13)}},
    legend columns=5,
    legend style={font=\scriptsize},
    legend image post style={scale=0.5},
    ymin=-30,
    ymax=5,
    y coord trafo/.code=\pgfmathparse{#1+30},
    y coord inv trafo/.code=\pgfmathparse{#1-30},
    xticklabels={0,-90,-60,-30,0,30,60,90,120,150,180,-150,-120},
    ]
    \addplot+[no marks,matlab1,thick]  table[x expr=\thisrowno{0}+90, y index=1] {data/MVDRPatternSOI0Int180_1870.txt};
    \addplot+[no marks,matlab2,thick]  table[x expr=\thisrowno{0}+90, y index=1] {data/MVDRPatternSOI0Int180_1350.txt};
    \addplot+[no marks,matlab3,thick]  table[x expr=\thisrowno{0}+90, y index=1] {data/MVDRPatternSOI0Int180_2400.txt};
    %addplot+[no marks, thick, red, dashed] coordinates {(-90, 5) (-90, -30) };
    %\addplot+[no marks, dashed, thick, black] coordinates {(-30, -30)};
    \draw[dashed, very thick, red] (axis cs:-90,-30) -- (axis cs:-90,5) node[above left, red, yshift=30pt]{Int.};
    \draw[dashed, very thick, blue] (axis cs:90,-30) -- (axis cs:90,5) node[below left, blue, yshift=-35pt]{SOI};
    \addlegendentry{$1.87\,\mathrm{GHz}$}
    \addlegendentry{$1.35\,\mathrm{GHz}$}
    \addlegendentry{$2.4\,\mathrm{GHz}$}

    \end{polaraxis}
    \end{tikzpicture}

%% file: figs/patternsinterferer90.tex
%\centering
    \begin{tikzpicture}
    \begin{polaraxis}[
    scale=0.8,
    ylabel style={xshift=0.1cm,yshift=-0.5cm},
    xlabel style={xshift=0.1cm,yshift=-0.5cm},
    legend style={anchor=north, at={(axis cs:270,13)}},
    legend columns=5,
    legend style={font=\scriptsize},
    legend image post style={scale=0.5},
    ymin=-30,
    ymax=5,
    y coord trafo/.code=\pgfmathparse{#1+30},
    y coord inv trafo/.code=\pgfmathparse{#1-30},
    xticklabels={0,-90,-60,-30,0,30,60,90,120,150,180,-150,-120},
    ]
    \addplot+[no marks,matlab1,thick]  table[x expr=\thisrowno{0}+90, y index=1] {data/MVDRPatternSOI0Int90_1870.txt};
    \addplot+[no marks,matlab2,thick]  table[x expr=\thisrowno{0}+90, y index=1] {data/MVDRPatternSOI0Int90_1350.txt};
    \addplot+[no marks,matlab3,thick]  table[x expr=\thisrowno{0}+90, y index=1] {data/MVDRPatternSOI0Int90_2400.txt};
    %addplot+[no marks, thick, red, dashed] coordinates {(-90, 5) (-90, -30) };
    %\addplot+[no marks, dashed, thick, black] coordinates {(-30, -30)};
    \draw[dashed, very thick, red] (axis cs:-180,-30) -- (axis cs:-180,5) node[above right, red, yshift=0pt]{Int.};
    \draw[dashed, very thick, blue] (axis cs:90,-30) -- (axis cs:90,5) node[below left, blue, yshift=-35pt]{SOI};
    \addlegendentry{$1.87\,\mathrm{GHz}$}
    \addlegendentry{$1.35\,\mathrm{GHz}$}
    \addlegendentry{$2.4\,\mathrm{GHz}$}

    \end{polaraxis}
    \end{tikzpicture}

%% file: figs/beamsteeringSINR1350_v2.tex
\begin{tikzpicture}
\begin{axis}[
legend style={font=\scriptsize},
legend columns=2,
legend pos=south west,
legend style={inner sep=2pt},
legend style={row sep=-2pt},
legend style={at={(0.5,-0.2)},anchor=north},
legend cell align={left},
legend image post style={scale=0.5},
small,
xlabel=Interferer angle $\mathrm{[^\circ]}$,
ylabel=SINR $\mathrm{[dB]}$,
width=1\textwidth,
height=1\textwidth,
xmin=-180,
xmax=180,
ymin=-15,
ymax=25,
grid=both,
thick,
ylabel near ticks,
ylabel shift = -0.2cm,
%axis y line*=left,
%ytick distance=10
%xtick distance=45
xtick={-135, -90, -45, 0, 45, 90, 135}
]
\addplot+[no marks,matlab1]  table[x=interfererangle,y=SINR] {data/MVDRSoI0_1350.txt};
\addplot+[no marks,matlab2]  table[x=interfererangle,y=SINR] {data/MVDRSoI46_1350.txt};

\addplot+[no marks,matlab1,dashed]  table[x=interfererangle,y=SINR] {data/PhaseOptimizedSoI0_1350.txt};
\addplot+[no marks,matlab2,dashed]  table[x=interfererangle,y=SINR] {data/PhaseOptimizedSoI46_1350.txt};

\addplot+[no marks,matlab3,dashed]  table[x=interfererangle,y=SINRantennaselect] {data/AntennaSelectionSoI0_1350.txt};
\addplot+[no marks,matlab4,dashed]  table[x=interfererangle,y=SINRantennaselect] {data/AntennaSelectionSoI46_1350.txt};

\draw[black, dashed, very thick] (axis cs:-180,18.5) -- (axis cs:180,18.5) node[above left]{No interference};
\draw[black, dashed, very thick] (axis cs:-180,-10) -- (axis cs:180,-10) node[below left]{No interference suppression};

%\addlegendentry{MVDR, 0$^\circ$ SOI}
%\addlegendentry{MVDR, 45$^\circ$ SOI}
%\addlegendentry{Phase-optimized, 0$^\circ$ SOI}
%\addlegendentry{Phase-optimized, 45$^\circ$ SOI}

\end{axis}
\end{tikzpicture}

%% file: figs/beamsteeringSINR_v2.tex
\begin{tikzpicture}
\begin{axis}[
legend style={font=\scriptsize},
legend columns=3,
legend pos=south west,
legend style={inner sep=2pt},
legend style={row sep=-2pt},
legend style={at={(0.5,-0.36)},anchor=north,overlay},
legend cell align={left},
legend image post style={scale=0.5},
small,
xlabel=Interferer angle $\mathrm{[^\circ]}$,
%ylabel=SINR $\mathrm{[dB]}$,
width=1\textwidth,
height=1\textwidth,
xmin=-180,
xmax=180,
ymin=-15,
ymax=25,
grid=both,
thick,
ylabel near ticks,
ylabel shift = -0.2cm,
%axis y line*=left,
%ytick distance=10
%xtick distance=45
xtick={-135, -90, -45, 0, 45, 90, 135}
]
\addplot+[no marks,matlab1]  table[x=interfererangle,y=SINR] {data/MVDRSoI0.txt};
\addlegendentry{MVDR, 0$^\circ$ SOI}

\addplot+[no marks,matlab1,dashed]  table[x=interfererangle,y=SINR] {data/PhaseOptimizedSoI0_1870.txt};
\addlegendentry{Phase-optimized, 0$^\circ$ SOI}

\addplot+[no marks,matlab3,dashed]  table[x=interfererangle,y=SINRantennaselect] {data/AntennaSelectionSoI0_1870.txt};
\addlegendentry{Antenna selection, 0$^\circ$ SOI}

\addplot+[no marks,matlab2]  table[x=interfererangle,y=SINR] {data/MVDRSoI46.txt};
\addlegendentry{MVDR, 45$^\circ$ SOI}

\addplot+[no marks,matlab2,dashed]  table[x=interfererangle,y=SINR] {data/PhaseOptimizedSoI46_1870.txt};
\addlegendentry{Phase-optimized, 45$^\circ$ SOI}

\addplot+[no marks,matlab4,dashed]  table[x=interfererangle,y=SINRantennaselect] {data/AntennaSelectionSoI46_1870.txt};
\addlegendentry{Antenna selection, 45$^\circ$ SOI}

\draw[black, dashed, very thick] (axis cs:-180,21) -- (axis cs:180,21) node[above left]{No interference};
\draw[black, dashed, very thick] (axis cs:-180,-10) -- (axis cs:180,-10) node[below left]{No interference suppression};

\end{axis}
\end{tikzpicture}

%% file: figs/beamsteeringSINR2400_v2.tex
\begin{tikzpicture}
\begin{axis}[
legend style={font=\scriptsize},
legend columns=2,
legend pos=south west,
legend style={inner sep=2pt},
legend style={row sep=-2pt},
legend style={at={(0.5,-0.2)},anchor=north},
legend cell align={left},
legend image post style={scale=0.5},
small,
xlabel=Interferer angle $\mathrm{[^\circ]}$,
%ylabel=SINR $\mathrm{[dB]}$,
width=1\textwidth,
height=1\textwidth,
xmin=-180,
xmax=180,
ymin=-15,
ymax=25,
grid=both,
thick,
ylabel near ticks,
ylabel shift = -0.2cm,
%axis y line*=left,
%ytick distance=10
%xtick distance=45
xtick={-135, -90, -45, 0, 45, 90, 135}
]
\addplot+[no marks,matlab1]  table[x=interfererangle,y=SINR] {data/MVDRSoI0_2400.txt};
\addplot+[no marks,matlab2]  table[x=interfererangle,y=SINR] {data/MVDRSoI46_2400.txt};

\addplot+[no marks,matlab1,dashed]  table[x=interfererangle,y=SINR] {data/PhaseOptimizedSoI0_2400.txt};
\addplot+[no marks,matlab2,dashed]  table[x=interfererangle,y=SINR] {data/PhaseOptimizedSoI46_2400.txt};

\addplot+[no marks,matlab3,dashed]  table[x=interfererangle,y=SINRantennaselect] {data/AntennaSelectionSoI0_2400.txt};
\addplot+[no marks,matlab4,dashed]  table[x=interfererangle,y=SINRantennaselect] {data/AntennaSelectionSoI46_2400.txt};

\draw[black, dashed, very thick] (axis cs:-180,21) -- (axis cs:180,21) node[above left]{No interference};
\draw[black, dashed, very thick] (axis cs:-180,-10) -- (axis cs:180,-10) node[below left]{No interference suppression};

%\addlegendentry{MVDR, 0$^\circ$ SOI}
%\addlegendentry{MVDR, 45$^\circ$ SOI}
%\addlegendentry{Phase-optimized, 0$^\circ$ SOI}
%\addlegendentry{Phase-optimized, 45$^\circ$ SOI}

\end{axis}
\end{tikzpicture}